\documentclass{revtex4}
\usepackage{amssymb,epsf}
\begin{document}

\title{Thermodynamics of a Kerr Newman de Sitter black hole}
\author{M. H. Dehghani$^{1,2}$}\email{dehghani@physics.susc.ac.ir}
\author{H. KhajehAzad$^{1}$}
\address{${1}$. Physics Department and Biruni Observatory, College of
Sciences,  Shiraz University, Shiraz 71454, Iran\\
        ${2}$. Institute for Studies in Theoretical Physics and Mathematics
        (IPM),\\
            P.O. Box 19395-5531, Tehran, Iran}

\begin{abstract}
We compute the conserved quantities of the four-dimensional
Kerr-Newman-dS (KNdS) black hole through the use of the
counterterm renormalization method, and obtain a generalized Smarr
formula for the mass as a function of the entropy, the angular
momentum and the electric charge. The first law of thermodynamics
associated to the cosmological horizon of KNdS is also
investigated. Using the minimal number of intrinsic boundary
counterterms, we consider the quasilocal thermodynamics of
asymptotic de Sitter Reissner-Nordstrom black hole, and find that
the temperature is equal to the product of the surface gravity
(divided by $2\pi$) and the Tolman redshift factor. We also
perform a quasilocal stability analysis by computing the
determinant of Hessian matrix of the energy with respect to its
thermodynamic variables in both the canonical and the
grand-canonical ensembles and obtain a complete set of phase
diagrams. We then turn to the quasilocal thermodynamics of
four-dimensional Kerr-Newman-de Sitter black hole for virtually
all possible values of the mass, the rotation and the charge
parameters that leave the quasilocal boundary inside the
cosmological event horizon, and perform a quasilocal stability
analysis of KNdS black hole.
\end{abstract}
\maketitle


\section{Introduction\label{Int}}

The observational evidence for a positive cosmological constant
and the proposed de Sitter conformal field theory (dS/CFT)
correspondence \cite{Str1} have provided motivations for studying
the thermodynamics of de Sitter spacetime in the presence of black
holes \cite{Cai}. Since asymptotic de Sitter black holes have a
cosmological event horizon, one may use a local description of
thermodynamics of the event or cosmological horizons separately
\cite{Pad}. Another way of studying the thermodynamics of these
kind of black holes is to investigate their quasilocal
thermodynamics \cite{Deh}.

One of the quantities associated with the gravitational
thermodynamics of black holes is the physical entropy $S$ which is
proportional to the area of the event horizon(s) \cite{Beck,Haw}.
The surprising and impressive features of the area law of entropy
is its universality which applies to all kinds of black holes and
black strings \cite{HHP,Mann}, and it also applies to the
cosmological event horizon of the asymptotic de Sitter black holes
\cite{GH1}. Another quantity of interest is the temperature $\beta
^{-1}$, which is proportional to the surface gravity of the event
horizon(s).

Other black hole properties such as energy, angular momentum and
conserved charges, can also be given a thermodynamic
interpretation \cite{GH2}. But as is known, these quantities
typically diverge for asymptotic flat, AdS, and dS spacetimes. A
common approach of evaluating them has been to carry out all
computations relative to some reference spacetime that is regarded
as the ground state for the class of spacetimes of interest
\cite{BY}. Unfortunately, it suffers from several drawbacks. The
choice of reference spacetime is not always unique \cite{CCM}, nor
is it always possible to embed a boundary with a given induced
metric into the reference background. Indeed, for rotating
spacetimes this latter problem creates a serious barrier against
calculating the subtraction energy, and calculations have only
been performed in the slow-rotating regime \cite{Mart}. An
extension of this approach that addresses these difficulties was
developed based on the conjectured AdS/CFT correspondence for
asymptotic AdS spacetimes \cite{Hen,BK,EJM}, and recently for
asymptotic de Sitter spacetimes \cite{Bal1,GM1,Deh1}. Indeed, the
near-boundary analysis of asymptotically AdS spacetimes can be
analytically continued to asymptotic dS spacetimes and therefore
the counterterms have a straightforward continuation. It is
believed that appending a counterterm, $I_{ct}$, to the action
which depends only on the intrinsic geometry of the boundary(ies)
can remove the divergences. This requirement, along with general
covariance, implies that these terms are functionals of curvature
invariants of the induced metric and have no dependence on the
extrinsic curvature of the boundary(ies). An algorithmic procedure
exists for constructing $I_{ct}$ for asymptotic AdS \cite{KLS} and
dS spacetimes \cite{GM1}, and so its determination is unique.
Addition of $I_{ct}$\ will not affect the bulk equations of
motion, thereby eliminating the need to embed the given geometry
in a reference spacetime. Hence thermodynamic and conserved
quantities can now be calculated intrinsic for any given
asymptotically AdS or dS spacetime. The efficiency of this
approach has been demonstrated in a broad range of examples for
both the asymptotic AdS and dS spacetimes
\cite{Bal1,DaM,GM1,Deh2,Deh3}.

Recently, we have considered the effects of including $I_{ct}$ for
quasilocal gravitational thermodynamics of Kerr, Kerr-AdS and
Kerr-dS black holes in which the region enclosed by $\partial
\mathcal{M}$ was spatially finite. We have also performed a
quasilocal stability analysis and found phase behavior that was
incommensurate with their previous analysis carried out at
infinity \cite{Deh,Deh4}. In this paper, we investigate the
thermodynamic properties of the class of Kerr-Newman-dS black
holes in the context of both temporally infinite and spatially
finite boundaries with the boundary action supplemented by
$I_{ct}$. With their lower degree of symmetry relative to
spherically symmetric black holes, these spacetimes allow for a
more detailed study of the consequences of including $I_{ct}$ for
temporally infinite and spatially finite boundaries. There are
several reasons for considering this. First, one can obtain the
thermodynamic quantities corresponding to an asymptotically de
Sitter spacetime within a fixed radius, while there is no
self-consistent thermodynamics for the whole spacetime.
Furthermore, the inclusion of $I_{ct}$ eliminates the embedding
problem from consideration, whether or not the temporary infinite
limit is taken, and so it is of interest to see what its impact is
on quasilocal thermodynamics.

The outline of our paper is as follows. We review the basic
formalism in Sec. \ref{Genfor}. In Sec. \ref{KN} we consider the
Kerr-Newman-dS$_4$ spacetime and compute the thermodynamic
quantities associated to the cosmological horizon. Also we give a
generalized Smarr formula and investigate the first law of
thermodynamics for the cosmological event horizon. In Sec.
\ref{RN}, we consider the quasilocal thermodynamics of
Reissner-Nordstrum-dS spacetime, and perform a stability analysis
through the use of the determinant of Hessian matrix of the energy
with respect to the entropy and the charge. In Sec. \ref{KNth}, we
compute the determinant of Hessian matrix of the energy with
respect to its thermodynamic variables for Kerr-Newman-de Sitter
black holes, numerically, and present our results graphically. We
finish our paper with some concluding remarks.

\section{General Formalism\label{Genfor}}

The gravitational action of four-dimensional asymptotically de Sitter
spacetimes $\mathcal{M}$, with boundary $\mathcal{\delta M}^{\pm }$ in the
presence of an electromagnetic field is
\begin{equation}
I_G=-\frac 1{16\pi }\int_{\mathcal{M}}d^4x\sqrt{-g}\left( \mathcal{R}%
-2\Lambda -F_{\mu \nu }F^{\mu \nu }\right) +\frac 1{8\pi }\int_{\mathcal{%
\delta M}^{-}}^{\mathcal{\delta M}^{+}}d^3x\sqrt{-\gamma }\Theta
(\gamma ) , \label{IG}
\end{equation}
where $F_{\mu \nu }=\partial _\mu A_\nu -\partial _\nu A_\mu $ is
the
electromagnetic tensor field and $A_\mu $ is the vector potential. In Eq. (%
\ref{IG}) $\mathcal{M}$ is the bulk manifold with metric $g_{\mu
\nu }$, and $\mathcal{\delta M}^{\pm }$ are spacial boundaries at
early and late times with induced metrics $\gamma _{ij}$ and the
extrinsic curvature $\Theta ^{\mu \nu }$. The first term is the
Einstein-Maxwell term with positive cosmological constant $\Lambda
=3/l^2$, and the second term is the Gibbons-Hawking boundary term
that is chosen such that the variational
principle is well defined. The notation $\int_{\mathcal{\delta M}^{-}}^{%
\mathcal{\delta M}^{+}}$ indicates an integral over the late-time
boundary minus the integral over the early-time boundary which are
both Euclidean denoted by $\mathcal{I}^{+}$ and $\mathcal{I}^{-}$
respectively. In general, $I_G$ of Eq. (\ref{IG}) is divergent
when evaluated on the solutions at $\mathcal{I}^{\pm}$, as is the
Hamiltonian and other associated conserved quantities. Rather than
eliminating these divergences by incorporating a reference term in
the spacetime, a new term $I_{ct}$ is added to the action that is
a functional only of boundary curvature invariants
\cite{Bal1,Deh1}. Although there may exist a very large number of
possible invariants, one could add only a finite number of them in
a given dimension. Quantities such as energy, mass, etc. can then
be understood as intrinsically defined for a given spacetime, as
opposed to being defined relative to some abstract (and
non-unique) background, although this latter option is still
available. In four dimensions the counterterm is \cite{Bal1}
\begin{equation}
I_{ct}=\frac 2l\frac 1{8\pi }\int_{\mathcal{\delta M}^{\pm }}d^3x\sqrt{%
-\gamma }\left( 1-\frac{l^2}4R\right) ,  \label{Ict}
\end{equation}
where $R$ is the Ricci scalar of the boundary metric $\gamma _{ij}$ and $%
\int_{\mathcal{\delta M}^{\pm }}$ indicates the sum of the
integral over the early and late time boundaries. Although other
counterterms (of higher mass dimension) may be added to $I_{ct}$,
they will make no contribution to the evaluation of the action or
Hamiltonian due to the rate at which they decrease toward
infinity, and we shall not consider them in our analysis here. It
is worthwhile to mentioning that the inclusion of matter fields in
the gravitational action requires the addition of further
counterterms when the dimension of the bulk is greater than four.
But since we consider the four-dimensional spacetimes, we don't
need any additional counterterms \cite {Sken2}.

A thorough discussion of the quasilocal formalism has been given
elsewhere \cite{BY} and so we only briefly recapitulate it here.
The total action can be written as a linear combination of the
gravity term (\ref{IG}) and the counterterms (\ref{Ict}) as
\begin{equation}
I=I_G+I_{ct} .  \label{Acta}
\end{equation}
Using the Brown and York definition \cite{BY} one can construct a
divergence-free stress-energy tensor from the total action
(\ref{Acta}) either on an early or late boundary as \cite{Bal1}:
\begin{eqnarray}
T^{+ab} &=&-\frac 1{8\pi }\left\{ (\Theta ^{ab}-\Theta \gamma ^{ab})-\frac
2l\left[ \gamma ^{ab}+\frac{l^2}2\left( R^{ab}-\frac 12R\gamma ^{ab}\right)
\right] \right\} ,  \label{Stres1} \\
T^{-ab} &=&-\frac 1{8\pi }\left\{ -(\Theta ^{ab}-\Theta \gamma ^{ab})-\frac
2l\left[ \gamma ^{ab}+\frac{l^2}2\left( R^{ab}-\frac 12R\gamma ^{ab}\right)
\right] \right\} .  \label{Stres2}
\end{eqnarray}
To obtain the boundary stress tensor, we should evaluate Eq. (\ref{Stres1})
at fixed time and send time to infinity. We will suppress the $+$ sign
notation in $T^{+ab}$ in the rest of the paper. To compute the thermodynamic
and conserved quantities of the spacetime, one should write the metric $%
\gamma _{ab}$ on equal time surfaces in the ADM form:
\[
\gamma _{ab}dx^adx^b=N^2d\rho ^2+\sigma _{ij}\left( d\phi
^i+V^id\rho \right) \left( d\phi ^j+V^jd\rho \right) ,
\]
where the coordinates $\phi ^i$ are the angular variables parameterizing the
closed surfaces around an origin. Then the total thermodynamic energy of the
system is \cite{BY}
\begin{equation}
E=\int_{\mathcal{B}}d^2\phi \sqrt{\sigma }\varepsilon;
\hspace{1cm} \varepsilon=T_{ab}u^au^b, \label{Entot}
\end{equation}
where $u^a$ is the unit normal vector on a surface of fixed $\rho $ and $%
\sigma $ is the determinant of the metric $\sigma _{ij}$. If the boundary of
the spacetime has an isometry generated by a Killing vector $\mathcal{\xi }%
^a $, then the conserved charge associated with $\mathcal{\xi }^a$
is \cite {Bal1}
\begin{equation}
\mathcal{Q}(\mathcal{\xi )}=\int_{\mathcal{B}}d^2\phi \sqrt{\sigma }T_{ab}u^a%
\mathcal{\xi }^b,  \label{charge}
\end{equation}
Since all the asymptotic de Sitter spacetimes have an asymptotic isometry
generated by $\partial /\partial t$, there is a notion of a conserved mass
as computed at future infinity as
\begin{equation}
M=\int_{\mathcal{B}}d^2\phi \sqrt{\sigma }T_{ab}u^a\xi ^b,  \label{Mastot}
\end{equation}
We compute the quantities of Eqs. (\ref{Mastot}) on a surface of
fixed time and then send time to infinity so that it approaches
the spacetime boundary at $\mathcal{I}^{\pm }$. Indeed, for $r$
greater than the radius of the cosmological horizon $t$ is a
spatial direction, while $r$ will become a temporal coordinate.
Thus, the large $r$ surfaces approach $\mathcal{I}^{\pm }$, and
since we purpose to define the stress tensor, and the mass at
$\mathcal{I}^{\pm }$, we will evaluate these quantities on the
surface of large $r$.

Similarly if the surface $\mathcal{B}$ has an isometry generated
by $\zeta =\partial /\partial \phi $, then the quantity
\begin{equation}
J=\int_{\mathcal{B}}d^2\phi \sqrt{\sigma }T_{ab}u^a\zeta ^b,
\label{Angtot}
\end{equation}
can be regarded as a conserved angular momentum associated with the surface $%
\mathcal{B}$. Many authors have used the counterterm method for
various asymptotic de Sitter spacetimes and computed the
associated conserved charges \cite{Bal1,GM1,Deh1}. We shall also
study the implications of including the counterterms (\ref{Ict})
for Reissner-Nordstrom-dS and Kerr-Newman-dS black holes in
spatially finite region.

\section{The Thermodynamics associated to the cosmological horizon of the
Kerr-Newman-dS$_4$ black hole\label{KN}}

Here we consider the charged rotating black holes in four
dimensions, whose general form is
\begin{eqnarray}
ds^2 &=&-\frac{\Delta _r^2}{\rho ^2}\left( dt-\frac a\Xi \sin ^2\theta d\phi
\right) ^2+\frac{\rho ^2}{\Delta _r^2}dr^2  \nonumber \\
&&\ +\frac{\rho ^2}{\Delta _\theta }d\theta ^2+\frac{\Delta _\theta \sin
^2\theta}{\rho ^2}\left( a dt-\frac{(r^2+a^2)}\Xi d\phi \right) ^2,
\label{met1a}
\end{eqnarray}
where
\begin{eqnarray}
\Delta _r^2 &=&(r^2+a^2)(1-r^2/l^2)-2mr+q^2,  \nonumber \\
\Delta _\theta &=&1+\frac{a^2}{l^2}\cos ^2\theta ,  \nonumber \\
\Xi &=&1+\frac{a^2}{l^2},  \nonumber \\
\rho ^2 &=&r^2+a^2\cos ^2\theta ,  \label{met1b}
\end{eqnarray}
and the vector potential $A_\mu $ is:
\begin{equation}
A_\mu =-\frac{qr}{\rho ^2}(\delta _\mu ^t-\frac{a\sin ^2\theta }\Xi \delta
_\mu ^\phi ).  \label{Pot}
\end{equation}
For $q=0$, the metric (\ref{met1a}) and (\ref{met1b}) is Kerr-dS$_4$
spacetime (or Kerr spacetime if $l\rightarrow \infty $), and for $a=0$ the
metric is Reissner Nordstrom-dS$_4$ spacetime, which has zero angular
momentum. Here the parameters $m$, $a$, and $q$ are associated with the
mass, the angular momentum and the charge of the spacetime, respectively.
The metric of Eqs. (\ref{met1a}) and (\ref{met1b}) has three horizons
located at $r_{-}$, $r_{+}$, and $r_c$, provided the parameters $m$, $l$, $a$%
, and $q$ are chosen suitably. The three parameters $l$, $a$ and $q$ should
be chosen such that $l^4+a^4-14a^2l^2-12q^2l^2\geq 0$, and $m$ should lie in
the range between $m_{1\hbox{\small{crit}}}\leq m\leq m_{2\hbox{\small{crit}}%
}$, where $m_{1\hbox{\small{crit}}}$ and $m_{2\hbox{\small{crit}}} $ are the
positive real solutions of the following equation:
\begin{eqnarray}
&& 27l^6m^4+(33l^4a^4-36l^6q^2-33l^6a^2+36l^4q^2a^2+l^2a^6-l^8)m^2
\nonumber \\
&&
+a^8q^2+22l^4a^4q^2+12l^6a^2q^2+6a^6l^4+4l^6a^4+32l^4a^2q^4+l^8q^2
\nonumber \\
&&
+12a^6l^2q^2+4l^2a^8+l^8a^2+a^{10}+16l^4q^6+8l^6q^4+8l^2q^4a^4=0.
\label{Crit}
\end{eqnarray}
It is worthwhile to mention that when $m=m_{1\hbox{\small{crit}}}$, $r_{-}$
and $r_{+}$ are equal and we have an extreme black hole. For the case in
which $m=m_{2\hbox{\small{crit}}}$, the radii of event and cosmological
horizons are equal and we have a critical hole. Also, one may note that in
the limit $l\rightarrow \infty $, $m_{1\hbox{\small{crit}}}=\sqrt{a^2+q^2}$,
$m_{2\hbox{\small{crit}}}\rightarrow \infty $, $r_{\pm }=m\pm \sqrt{%
m^2-a^2-q^2}$, and $r_c\rightarrow \infty $.

For given values of $a$ and $m$, the parameter $l$ should be greater than $%
l_{\hbox{\small{crit}}}$, where $l_{\hbox{\small{crit}}}$ is the
positive real solution of Eq. (\ref{Crit}). When
$l=l_{\hbox{\small{crit}}}$, we have a critical hole. Also the
positive real solution of Eq. (\ref{Crit}) for $a$ gives the
critical value of $a$ (upper bound for $a$) for which the hole is
extremal. The two real solutions of Eq. (\ref{Crit}) for $q$
determine the lower and upper bounds of $q$ that belong to the
critical and extremal black hole, respectively.

Since the temperature associated with the event or cosmological
horizons has different values for asymptotic de Sitter spacetimes
one should use a local description of the thermodynamics of the
black hole. In the remain of this section we investigate the
thermodynamics associated to the cosmological horizon and in the
next section we investigate the thermodynamic quantities
associated with the spacetime within an arbitrary surface with
fixed radius $r$.

Analytical continuation of the Lorentzian metric by $%
t\rightarrow i\tau $ and $a\rightarrow ia$ yields the Euclidean
section, whose regularity at $r=r_c$ requires that we should
identify $(\tau,\phi) \sim (\tau
+\beta _c, \phi +i\beta _c\Omega_c) $, where $\beta _c$ and $%
\Omega_c $ are the inverse Hawking temperature and the angular
velocity of the cosmological event horizon given as \cite{GH1}:
\begin{eqnarray}
&&\beta _c=\frac{4\pi l^2r_c(r_c^2+a^2)}{%
3r_c^4+a^2r_c^2-l^2r_c^2+a^2l^2+q^2l^2},  \label{betc} \\
&&\Omega _c=\frac{a\Xi }{r_c^2+a^2}.  \label{Ang}
\end{eqnarray}
Since the area law of the entropy is universal, and it can be used for the
cosmological horizon of dS black holes \cite{GH1}, the entropy is
\begin{equation}
S=\frac{\mathcal{A}_c}4=\frac{\pi (r_c^2+a^2)}\Xi .  \label{Sc}
\end{equation}

As we discussed in Sec. \ref{Genfor}, one can compute the finite
action through the use of boundary counterterms. To do this one
should take a space-like hypersurface, $\mathcal{P}$ of large
constant $r$ ($r_c<r<\infty$) in the future and calculate the
total action (\ref{Acta}). In the limit that $r$ tends to
infinity, this surface tends to future infinity $\mathcal{I}^+$ as
expected. The first integral of the action in (\ref{IG}) should be
first computed from $r_c$ to $r$, and then the total action should
be integrated on the hypersurface $\mathcal{P}$. Sending $r$ to
infinity the total action of the system is :
\begin{equation}
I =-\beta
_c\frac{(r_c^2+a^2)(r_c^3+r_ca^2+ml^2)-r_cl^2q^2}{2l^2\Xi
(r_c^2+a^2)},  \label{Ia}
\end{equation}
where $r_c$ and $\beta _c$ are the radius and the inverse of the
Hawking temperature of the cosmological event horizon. The total
mass can be calculated through the use of Eq. (\ref{Mastot}) and
sending the temporal coordinate to infinity. We obtain
\begin{equation}
M=-\frac m{\Xi ^2}.  \label{M}
\end{equation}
It is worthwhile to note that as $a$ and $q$ go to zero, the
action and mass given by Eqs. (\ref{Ia}) and (\ref{M}) reduce to
those of Schwarzschild-dS black hole obtained in \cite{Bal1}. The
total angular momentum can be computed through the use of Eq.
(\ref{Angtot}) for a boundary with finite radius, $r_+<r<r_c$. It
is a matter of calculation to show that $J$ is independent of $r$
given as:
\begin{equation}
J=-\frac{ma}{\Xi ^2}.  \label{J}
\end{equation}

We now obtain the mass as a function of the extensive quantities
$S$, $J$ and $Q$, where $Q$ defined as $q/\Xi $. Using the
expressions (\ref{M}) and (\ref{Sc}) for the mass, the angular
momentum and the entropy, and the fact that $\Delta_r(r_{c})=0$,
one can obtain a generalized Smarr formula as
\begin{equation}
M=-\left\{ \frac S{4\pi }+\frac \pi {4S}(4J^2+Q^4)+\frac{Q^2}2-\frac{J^2}{l^2%
}-\frac S{2\pi l^2}\left( Q^2+\frac S\pi -\frac{S^2}{2\pi
l^2}\right) \right\}^{1/2} .  \label{Smar}
\end{equation}
It is worthwhile to mentioning that the analytic continuation
$l\rightarrow il$ (Wick rotates) changes the above Smarr formula
to the Smarr formula of the Kerr-Newman-AdS black hole given in
Ref. \cite{Cal} Now, one may regard the parameters $S$, $J$ and
$Q$ as a complete set of extensive parameters for the mass
$M(S,J,Q)$ and define the intensive parameters conjugate to $S$,
$J$ and $Q$. These quantities are the temperature, the angular
velocities, and the electric potential defined as
\begin{equation}
T=\left( \frac{\partial M}{\partial S}\right) _{J,Q},\ \ \Omega =\left(
\frac{\partial M}{\partial J}\right) _{S,Q},\ \ \Phi =\left( \frac{\partial M%
}{\partial Q}\right) _{S,J}.  \label{Dsmar}
\end{equation}
It is a matter of straightforward calculation to show that the
temperature calculated from Eq. (\ref{Dsmar}) coincides with Eq.
(\ref{betc}), and the angular velocity obtained from Eq.
(\ref{Dsmar}) is the thermodynamic angular velocity that is equal
to $\Omega_c-a/l^2$. Also one can obtain the electric potential
$\Phi$ as
\begin{equation}
\Phi =-\frac{qr_c}{r_c^2+a^2}.  \label{Ang}
\end{equation}
It is worthwhile to note that the thermodynamic quantities calculated in
this section satisfy the first law of thermodynamics,
\begin{equation}
dM=TdS+\Omega dJ+\Phi dQ .  \label{Flth}
\end{equation}

\section{The quasilocal thermodynamics of a Reissner-Nordstrom-dS$_4$ black
hole \label{RN}}

First we investigate the quasilocal thermodynamics of a
non-rotating case in which all the thermodynamic quantities can be
integrated easily. For $a=0$, one obtains the quasilocal energy of
a Reissner-Nordstrom-dS black hole as
\begin{equation}
E=\frac l2\left( 1-2\frac{r^2}{l^2}\right) -r\sqrt{1-\frac{r^2}{l^2}-\frac{2m%
}r+\frac{q^2}{r^2}} \ .  \label{Ern}
\end{equation}
Using the expressions $Q=q$ for the charge and $S=\pi
(r_{+}^2+a^2)/4$ for the entropy of the event horizon, one can
write $E$ as a function of thermodynamic quantities $S$ and $Q$.
Now identifying the energy (\ref{Ern}) as the thermodynamic
internal energy for the Reissner-Nordstrom-dS black hole within
the boundary $r$, then the corresponding temperature and
electrostatic potential at the boundary with fixed radius
$r_+<r<r_c$ are
\begin{eqnarray}
&& T= \left(\frac{\partial E}{\partial S}\right)_Q=\frac 1{\sqrt{1-\frac{r^2}{l^2}-\frac{2m}r+%
\frac{q^2}{r^2}}}\left( \frac{1-3\frac{r_{+}^2}{l^2}+\frac{q^2}{r_{+}^2}}{%
4\pi r_{+}}\right) ,  \label{TRN}\\
&& \Phi= \left(\frac{\partial E}{\partial Q}\right)_S=\frac 1{\sqrt{1-\frac{r^2}{l^2}-\frac{2m}r+%
\frac{q^2}{r^2}}}\left( \frac{q}{r_+}-\frac{q}{r}\right) .
\label{PRN}
\end{eqnarray}
The first factor in the above expressions is the inverse of the
lapse function for Reissner-Nordstrom-dS$_4$ metric evaluated at
$r$, and is the Tolman redshift factor for temperature in a static
gravitational field \cite{Tol}. The second factors in Eq.
(\ref{TRN}) and (\ref{PRN})are the surface gravity ($\kappa _H$)
of the black hole divided by $2\pi $, and the electrostatic
potential of the hole relative to electrostatic potential at $r$.
Therefore as in the case of Schwarzschild-AdS and
Reissner-Nordstrom-AdS black holes \cite{Crei,Pec} the temperature
corresponding to the spacetime within the surface of fixed $r$ is
the product of $\kappa _H/2\pi $ and the redshift factor. Note
that the term due to the counterterm does not affect the
temperature and the electrostatic potential of the black hole.
\begin{figure}[tbp]
\epsfxsize=6cm \centerline{\epsffile{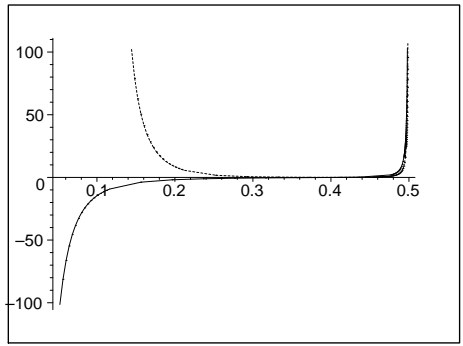}}
\caption{$(\partial ^2E/\partial S^2)_Q$ versus $r_{+}$ for $l=1$, $a=0$, $%
q=0.2$, (solid), and $0$ (dotted).} \label{Figure1}
\end{figure}
\begin{figure}[tbp]
\epsfxsize=6cm \centerline{\epsffile{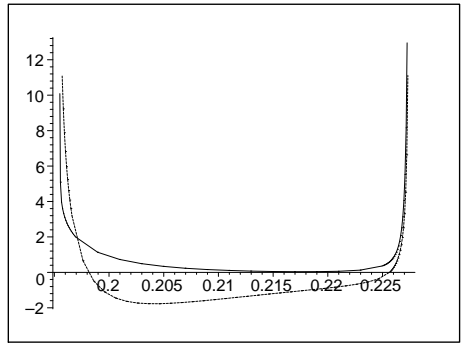}}
\caption{Determinant of Hessian matrix (solid) and $(\partial
^2E/\partial S^2)_Q$ (dashed) versus $m$ for $l=1$, $a=0$ and
$q=0.2$.} \label{Figure2}
\end{figure}

\subsection{Stability analysis in the canonical and the grand canonical ensemble}

The local stability analysis in any ensemble can, in principle, be
carried out by finding the determinant of the Hessian matrix of
$S$ with respect to its thermodynamic variables,
$\mathbf{H}_{X_iX_j}^S=|\partial ^2S/\partial X_i\partial X_j|$
where $X_i$'s are the thermodynamic variables of the system.
Indeed, the system is locally stable if the determinant of the
Hessian matrix satisfies $\mathbf{H}_{X_i,X_j}^S\leq 0$
\cite{Cev}. Also, one can perform the stability analysis through
the use of the determinant of Hessian matrix of the energy with
respect to its thermodynamic variables, and the stability
requirement $\mathbf{H}_{X_i,X_j}^S\leq 0$ may be rephrased as
$\mathbf{H}_{Y_i,Y_j}^E\geq 0$ \cite{Gub}.

The number of the thermodynamic variables depends on the ensemble
which is used. In the canonical ensemble, the charge is a fixed
parameter, and therefore the positivity of the thermal capacity
$(\partial ^2E/\partial S^2)_Q$ is sufficient to assure the local
stability. It is a matter of calculation to show that $(\partial
^2E/\partial ^2S)_Q$ is
\begin{eqnarray}
\frac{\partial ^2E}{\partial S^2}=\frac r{16\pi ^2l^4r_{+}^6\Delta _0^3}
&[&r(3r_{+}^4l^4-2r_{+}^6l^2-6r_{+}^2l^4q^2+3r_{+}^8+18r_{+}^4q^2l^2-5q^4l^4)
\nonumber \\
&&\ -2r_{+}(3r_{+}^4-3l^2q^2+r_{+}^2l^2)(l^2r^2-r^4+q^2l^2)],  \label{dERN}
\end{eqnarray}
where $\Delta _0$ is $\Delta _r$ at $a=0$. Note that $\Delta _0$
is equal to zero at horizons, and therefore $(\partial
^2E/\partial ^2S)_Q$ goes to infinity as $r$ approaches the radius
of event or cosmological horizon. we set the radius of the
boundary equal to $r=0.5$. For a given $l$ and a fixed value of
$r$ in the range $r_{+}<r<r_c$, Eq. (\ref{dERN}) has two real
solutions for $r_{+}$, which shows that there exist two stable
phases separated by an unstable intermediate mass phase. This is
not true for the case of an uncharged hole for which Eq.
(\ref{dERN}) has a single real solution (see Fig. \ref{Figure1}).

In the grand-canonical ensemble, the stability analysis can be
carried out by calculating the determinant of Hessian matrix of
the energy with respect to $S$ and $Q$,
\begin{eqnarray}
\mathbf{H}_{S,Q}^E &=&\frac{r^2}{16\pi ^2r_{+}^5(r-r_{+})}%
\{3r_{+}^8+q^4l^4-2l^2r_{+}^6+3l^4r_{+}^4+4rr_{+}^3q^2l^2  \nonumber \\
&& \hspace{1.5cm}
+2r_{+}(l^2-r^2)(rl^2q^2-2r_{+}l^2q^2-rl^2r_{+}^2-3rr_{+}^4)\}  \nonumber \\
&&\hspace{3cm} \times
\{rr_{+}{}^3+r^2r_{+}{}^2-r_{+}l^2r+r_{+}r^3+q^2l^2\}^{-2}.  \label{HRN}
\end{eqnarray}
Again note that the determinant of Hessian matrix goes to infinity
as $r$ goes to the radius of the event or cosmological horizon.
Since the number of thermodynamic variables in the canonical
ensemble is more than that of the grand-canonical ensemble, the
region of stability is smaller for the latter case. This fact can
be seen in Fig. \ref{Figure2}, which displays $(\partial
^2E/\partial ^2S)_Q$ and the determinant of the Hessian matrix as
a function of the the mass parameter $m$.

\section{The stability analysis of the Kerr-Newman-dS$_4$ black hole%
\label{KNth}}

Now we turn to the quasilocal thermodynamics of Kerr-Newman-dS$_4$
black hole. The energy of the system can be calculated through the
use of Eq. (\ref{Entot}) where the boundary $\mathcal{B}$ is a two
dimensional surface with fixed radial coordinate $r_+<r<r_c$. This
energy can be identify as the thermodynamic internal energy for
the Kerr-Newman-dS spacetime within the boundary $r$.

It is important to say that the total angular momentum calculated
by Eq. (\ref {Angtot}) does not depend on the radius of the
boundary surface and it is equal to the value given in Eq.
(\ref{J}). Now using Eq. (\ref{J}) for the angular momentum, and
the expressions $Q=q/\Xi $ and $S=\pi (r_{+}^2+a^2)/4$ for the
charge and the entropy, one can obtain the metric parameters in
terms of $S$, $J$ and $Q$:
\begin{eqnarray}
a &=&2\Gamma ^{-1/2}\sqrt{S\pi ^3}l^2J,  \nonumber \\
q &=&Q(1+4\pi ^3J^2l^2S\Gamma ^{-1}),  \nonumber \\
m &=&\frac 1{2l^2\pi ^{3/2}S^{1/2}\Gamma }(1+4\pi ^3J^2l^2S\Gamma ^{-1})^2,
\label{amq}
\end{eqnarray}
where
\begin{equation}
\Gamma \equiv S^4-4\pi ^3J^2l^2S+4\pi ^4J^2l^4+\pi ^4Q^4l^4+2\pi
^3Q^2l^4S-2\pi ^2Q^2l^2S^2+S^2l^4\pi ^2-2S^3l^2\pi.  \label{gam}
\end{equation}
With these expressions for $a$, $m$ and $q$, we can write $E$ as a
function of thermodynamic quantities $S$, $J$ and $Q$, and regard
$E(S,J,Q)$ as the thermodynamic internal energy for the
Kerr-Newman-dS black hole within the boundary with radius $r$. In
the canonical ensemble, the charge and angular momentum are fixed
parameters, and for this reason the positivity of the $(\partial
^2E/\partial S^2)_{J,Q}$ is sufficient to assure the local
stability. For general values of the metric parameters, however,
we cannot analytically calculate the internal energy as a function
of $S$, $J$ and $Q$, and so we resort to numerical integration. To
do this, we find $\varepsilon$ in Eq. (\ref{Entot})as a function
of $S$, $J$ and $Q$, calculate the derivative of
$\varepsilon(S,J,Q)$ and then compute the integration numerically.
We set the radius of the boundary equal to $r=0.56$, and we apply
the restrictions discussed in Sec. \ref{KN} on the parameters $l$,
$m$, $a$ and $q$. Note that for $r=0.56$, each of these parameters
has two critical values, one due to the extreme Kerr-Newman-dS
black hole and the other due to the case in which this radius is
equal to the radii of event and cosmological horizons.

\begin{figure}[tbp]
\epsfxsize=6cm \centerline{\epsffile{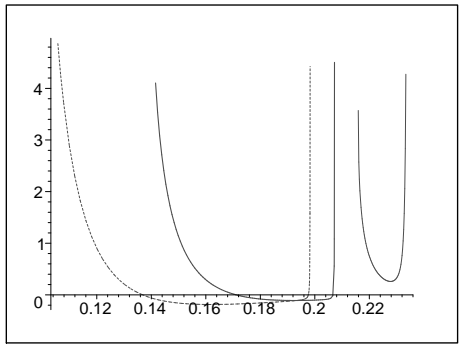}}
\caption{$(\partial ^2E/\partial S^2)_{J,Q}$ versus $m$ for $l=1$, $a=0.1$, $%
q=0$ (dotted), $0.1$ (thick-line), and $0.2$ (solid).}
\label{Figure3}
\end{figure}

\begin{figure}[tbp]
\epsfxsize=6cm \centerline{\epsffile{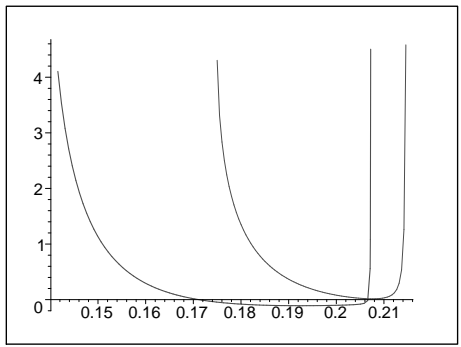}}
\caption{$(\partial ^2E/\partial S^2)_{J,Q}$ versus $m$ for $l=1$, $q=0.1$, $%
a=0.1$ (thick-line), and $0.15$ (solid).} \label{Figure4}
\end{figure}

\begin{figure}[tbp]
\epsfxsize=6cm \centerline{\epsffile{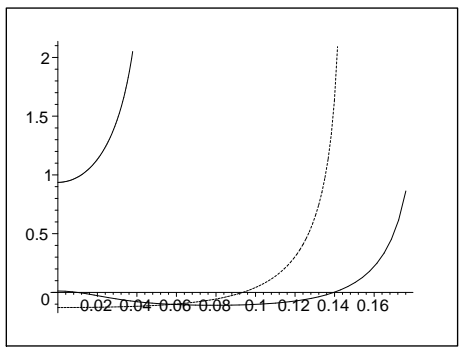}}
\caption{$(\partial ^2E/\partial S^2)_{J,Q}$ versus $a$ for $l=1$, $m=0.2$, $%
q=0.2$ (thick-line), $0.15$ (dotted), and $0.1$ (solid).}
\label{Figure5}
\end{figure}

\begin{figure}[tbp]
\epsfxsize=6cm \centerline{\epsffile{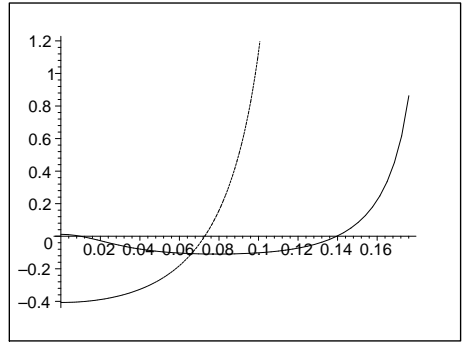}}
\caption{$(\partial ^2E/\partial S^2)_{J,Q}$ versus $a$ for $l=1$, $q=0.1$, $%
m=0.15$ (dotted), and $0.2$ (solid).} \label{Figure6}
\end{figure}

\begin{figure}[tbp]
\epsfxsize=6cm \centerline{\epsffile{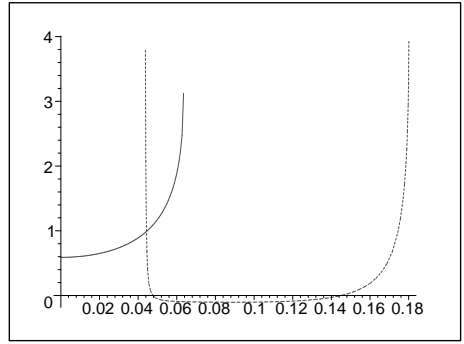}}
\caption{$(\partial ^2E/\partial S^2)_{J,Q}$ versus $q$ for $l=1$, $m=0.2$, $%
a=0.2$ (solid), and $0.1$ (dotted).} \label{Figure7}
\end{figure}

\begin{figure}[tbp]
\epsfxsize=6cm \centerline{\epsffile{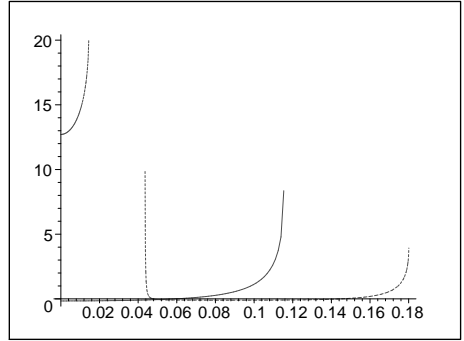}}
\caption{$(\partial ^2E/\partial S^2)_{J,Q}$ versus $q$ for $l=1$, $a=0.1$, $%
m=0.1$ (solid), $0.15$ (thick-line), and $0.2$ (dotted).}
\label{Figure8}
\end{figure}

Figure \ref{Figure3} displays the $m$ dependence of $(\partial
^2E/\partial S^2)_{J,Q}$ and shows that for a given value of $a$
and a large value of $q$, only a stable phase exists for all the
allowed values of $m$ ($m_{1\hbox{\small{crit}}}\leq m\leq
m_{\hbox{2\small{crit}}}$), but as $q$ decreases there begins to
appear an unstable phase between the two stable phases. Also Fig.
\ref{Figure4} shows the $m$ dependence of $(\partial ^2E/\partial
S^2)_{J,Q}$ for small $q$ and different values of $a$. It points
out that there exists an unstable phase, between two stable
phases, which begins to disappear as $a$ increases. Figures
\ref{Figure5} and \ref{Figure6} display the $a$ dependence of
$(\partial ^2E/\partial S^2)_{J,Q}$. They show that for large
values of $m$ and $q$ the hole is stable while for large $m$ and
small $q$, there exist two stable phases separated by an unstable
intermediate angular momentum phase. Also note that for small and
intermediate $q$ or $m$, the low angular momentum phase is not
stable (see the dotted curve in Fig. \ref {Figure5} and \ref
{Figure6}). Figure \ref{Figure7} which displays the $q$ dependence
of $(\partial ^2E/\partial S^2)_{J,Q}$ shows that for a large
value of $m$ and $a$ there exists only a stable phase, but an
unstable phase begin to appear as $a$ decreases. We also find that
for small $a$ and intermediate $m$ the black holes with small $q$
are not stable (see Fig.\ref{Figure8}).
It is worthwhile to note that in all of these cases $%
(\partial ^2E/\partial S^2)_{J,Q}$ goes to infinity as the hole
approaches its extreme case and therefore the nearly extreme black
hole is stable in the canonical ensemble. This stability analysis
is in qualitative agreement with that of Davies \cite{Davies}, who
performed a thorough analysis of the thermodynamic properties of
asymptotic Kerr-Newman-dS black holes. However, Davies used the
ADM mass parameters, whereas we are considering the quasilocal
energy $E$ in Eq. (\ref{Entot}) as a function of $r$, $S$, $J$ and
$Q$.

\begin{figure}[tbp]
\epsfxsize=6cm \centerline{\epsffile{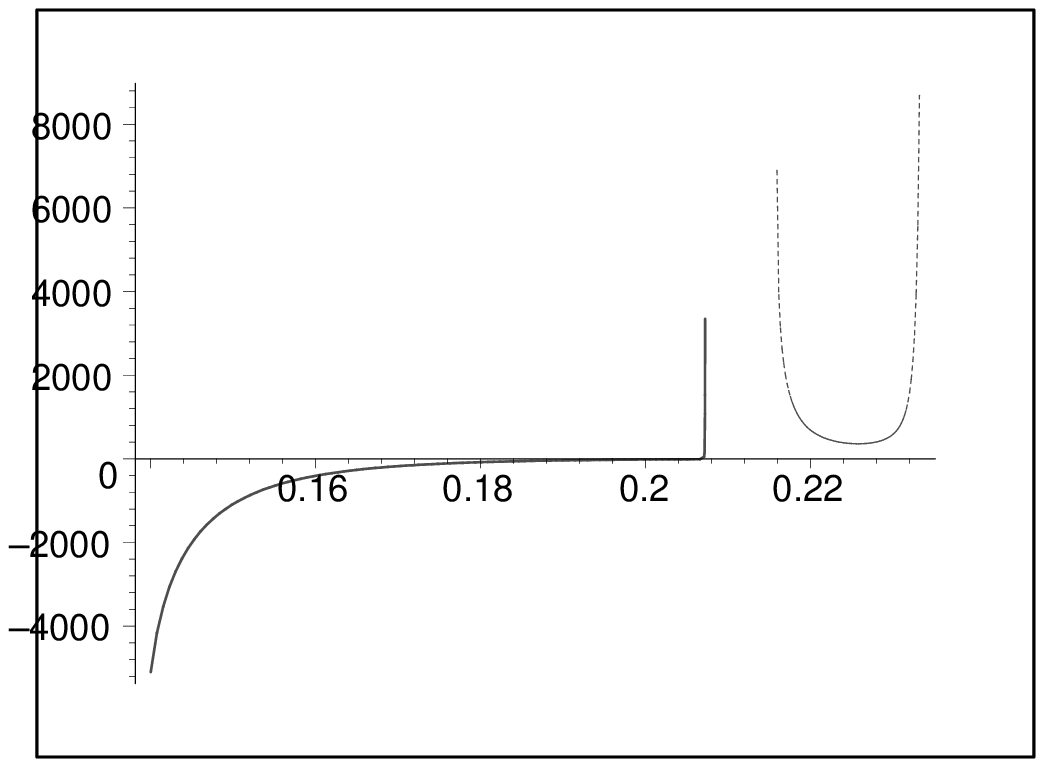}}
\caption{Determinant of Hessian matrix versus $m$ for $l=1$,
$a=0.1$, $q=0.1$ (thick-line), and $0.2$ (dashed).}
\label{Figure9}
\end{figure}
\begin{figure}[tbp]
\epsfxsize=6cm \centerline{\epsffile{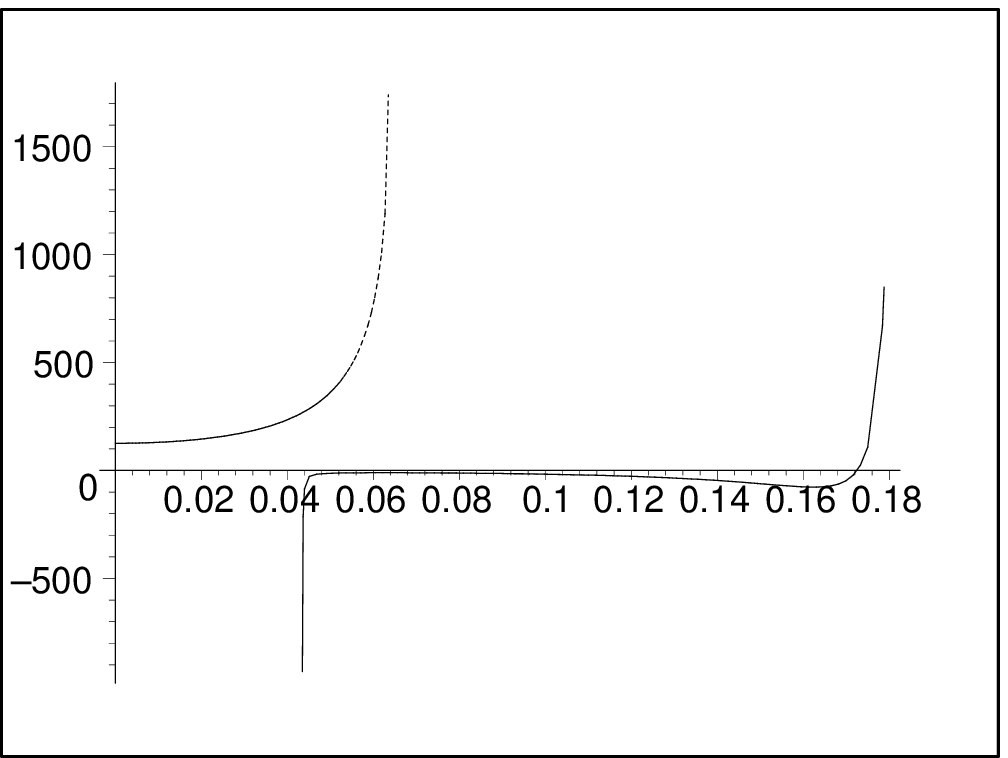}}
\caption{Determinant of Hessian matrix versus $q$ for $l=1$,
$m=0.2$, $a=0.2$ (dotted), and $0.1$ (solid).} \label{Figure10}
\end{figure}
\begin{figure}[tbp]
\epsfxsize=6cm \centerline{\epsffile{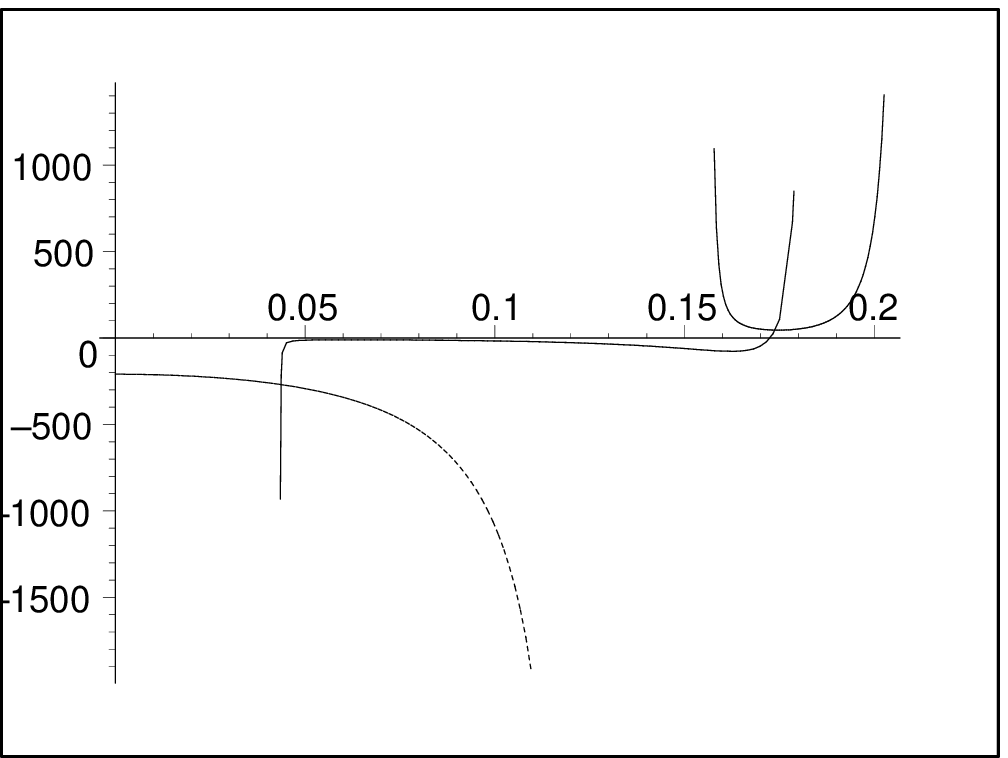}}
\caption{Determinant of Hessian matrix versus $q$ for $l=1$, $a=0.1$, $%
m=0.15 $ (dashed), $0.2$ (thick-line), and $0.22$ (solid).}
\label{Figure11}
\end{figure}
\begin{figure}[tbp]
\epsfxsize=6cm \centerline{\epsffile{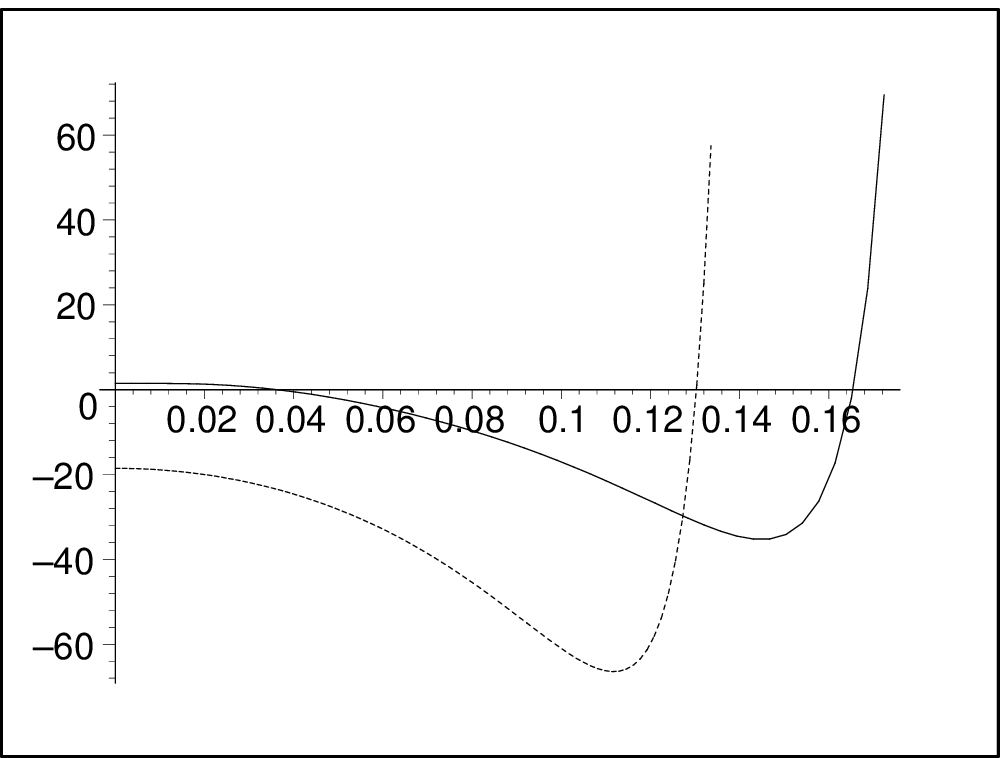}}
\caption{Determinant of Hessian matrix versus $a$ for $l=1$, $m=0.2$, $%
q=0.15 $ (dotted), and $0.1$ (solid).} \label{Figure12}
\end{figure}

In the grand-canonical ensemble, the thermodynamic variables are
the entropy, the charge and the angular momentum. The determinant
of the Hessian matrix can be calculated numerically. Some typical
results of these calculations are displayed in Figs.
\ref{Figure9}-\ref{Figure12}. As one can see from Fig.
\ref{Figure9} which shows the $m$ dependence of the
determinant of Hessian matrix, for a given value of $a$ and large values of $%
q$, the hole is stable, but for small values of $q$, only the
extreme case is stable. The $q$ dependence of the determinant of
the Hessian matrix are shown in Figs. \ref{Figure10} and
\ref{Figure11}. Figure \ref{Figure10} shows that for large values
of $m$ and $a$ we encounter a single stable phase, while for large
$m$ and small $a$ the only stable phase is the nearly extremal
hole. We also find that for small $a$ the hole is stable for large
$m$ and unstable for small $m $. For intermediate $m$ only the
nearly extreme hole is stable (see Fig. \ref{Figure11}). The $a$
dependence of the determinant of the Hessian matrix (Fig.
\ref{Figure12}) shows that for large values of $m$, when $q$ is
large the only stable phase is the nearly extreme hole, but as $q$
decreases a stable low angular momentum phase will appear. The
comparison of the figures of the canonical and grand-canonical
ensemble shows that the region of stability is smaller in the
latter case as stated before, and there are cases that even the
extreme hole is not stable in the grand-canonical ensemble.\\

\section{Closing Remarks\label{Clos}}

In this paper, we used the counterterm method to calculate the
conserved quantities and the Euclidean action of the
four-dimensional Kerr-Newman-dS black hole. Also we obtained a
generalized Smarr formula for the mass as a function of the
entropy, the angular momentum and the electric charge and showed
that these quantities satisfy the first law of thermodynamics
associated to the cosmological horizon. Since there is no
well-defined thermodynamics for the asymptotically de Sitter black
holes, a local investigation of the thermodynamics is necessary.
Our investigation of the boundary-induced counterterm prescription
(\ref{Ict}) at finite distances has allowed us to obtain a number
of interesting results, which we recapitulate here.

We were able to compute the quasilocal energy, the temperature, and the
Hessian matrix of the energy with respect to $S$ and $Q$ for arbitrary
values of the parameters of the Reissner-Nordstom-dS black hole
analytically. We found that the temperature at fixed $r$ is the product of $%
\kappa _H/2\pi $ and the redshift factor. We perform a stability analysis
both in the canonical and grand-canonical ensemble, and obtained a complete
phase diagrams. We found that at a fixed value of $r$ there exists two
stable phases separated by an unstable phase with intermediate mass value.
That is, charge causes the low mass black hole to be stable, a fact which is
not true for the Schwarzschild-dS black holes.

We also computed the determinant of the Hessian matrix,
quasilocally, for arbitrary values of the parameters of the
Kerr-Newman-dS$_4 $ solution, apart from the mild reality
restrictions considered in Sec. \ref {KN}. These quasilocal
quantities are intrinsically calculable numerically at fixed $r$
without any reference to a background spacetime. By finding the
parameters $a$, $m$ and $q$ in terms of the thermodynamic
quantities $S$, $J$ and $Q$, we regarded the interior of the
quasilocal surface as a thermodynamic system whose energy $E$ is a
function of $S$, $J$ and $Q$. We studied the stability of the
Kerr-Newman-dS black hole both in the canonical and
grand-canonical ensembles. Thereby we encountered an interesting
phase structure. We found that in the canonical ensemble the
extreme black hole is stable for all the allowed values of the
metric parameters while this is not true in the grand-canonical
ensemble. The stability analysis in the canonical ensemble yielded
results that are in qualitative agreement with previous
investigations carried out by using the ADM mass parameter
\cite{Davies}. This stability analysis is, of course local, and
phase transitions to other spacetimes of lower free energy might
exist.


\begin{thebibliography}{99}
\bibitem{Str1}  A. Strominger, J. High Energy Phys. \textbf{10}, 034 (2001);
\textbf{11}, 049 (2001); V. Balasubramanian, P. Horova, and D. Minic, \emph{%
ibid.} \textbf{05}, 043 (2001); E. Witten, hep-th/0106109; D.
Mu-In Park,
Phys. Lett. \textbf{B440}, 275 (1998); S. Nojiri and S. D. Odintsov, \emph{%
ibid}. \textbf{519}, 145 (2001); J. High Energy Phys. \textbf{12},
033 (2001); Phys. Lett. B \textbf{523}, 165 (2001); \textbf{528},
169 (2002); T. Shiromizu, D. Ida, and T. Torii, J. High Energy
Phys. \textbf{11}, 010 (2001); R. G. Cai, Phys. Lett. B
\textbf{525}, 331 (2002); Nucl. Phys. \textbf{B628}, 375 (2002);
R. G. Cai, Y. S. Myung, and Y. Z. Zhang, Phys. Rev. D \textbf{65},
084019 (2002); R. Bousso, A. Maloney, and A. Strominger,
\emph{ibid.} \textbf{65}, 104039 (2002); M. Spradlin and A. Volovich, \emph{%
ibid.} \textbf{65}, 104037 (2002); A. J. M. Medved,
hep-th/0205251; V. Balasubramanian, J. D. Boer and D. Minic,
hep-th/0207245; K. Skenderis, Class. Quantum Grav. \textbf{19},
5649 (2002).

\bibitem{Cai}  R. G. Cai, Phys. Lett. B \textbf{525}, 331 (2002);
hep-th/0112253; M. Cvetic, S. Nojiri and S. D. Odintsov, Nucl. Phys. \textbf{%
B628}, 295 (2002).

\bibitem{Pad}  T. Padmanabhan, Mod. Phys. Lett. \textbf{A17}, 923 (2002);
Class. Quantum Grav. \textbf{19} 5387 (2002).

\bibitem{Deh}  M. H. Dehghani, Phys. Rev. D \textbf{65}, 104030 (2002).

\bibitem{Beck}  J. D. Beckenstein, Phys. Rev. D \textbf{7}, 2333 (1973).

\bibitem{Haw}  S. W. Hawking, Nature \textbf{248}, 30 (1974) ; Comm. Math.
Phys. \textbf{43}, 199 (1975).

\bibitem{HHP}  S. W. Hawking and C. J. Hunter, Phys. Rev. D \textbf{59},
044025 (1999); C. J. Hunter, \emph{ibid.} \textbf{59}, 024009
(1999); S. W. Hawking, C. J. Hunter and D. N. Page, \emph{ibid}.
\textbf{59}, 044033 (1999).

\bibitem{Mann}  R. B. Mann, Phys. Rev. D \textbf{60}, 104047 (1999); \emph{%
ibid} \textbf{61}, 084013 (2000).

\bibitem{GH1}  G. W. Gibbons and S. W. Hawking, Phys. Rev. D \textbf{15},
2738 (1977).

\bibitem{GH2}  G. W. Gibbons and S. W. Hawking, Commun. Math Phys. \textbf{66%
}, 291 (1979).

\bibitem{BY}  J. D. Brown and J. W. York, Phys. Rev. D \textbf{47}, 1407
(1993).

\bibitem{CCM}  K. C. K. Chan, J. D. E. Creighton and R. B. Mann, Phys. Rev.
D \textbf{54}, 3892 (1996).

\bibitem{Mart}  E. A. Martinez, Phys. Rev. D \textbf{50}, 4920 (1994).

\bibitem{Hen}  M. Hennigson and K. Skenderis, J. High Energy Phys. \textbf{7}%
, 023 (1998).

\bibitem{BK}  V. Balasubramanian and P. Kraus, Commun. Math. Phys. \textbf{%
208,} 413 (1999).

\bibitem{EJM}  R. Emparan, C.V. Johnson and R. C. Myers, Phys. Rev. D
\textbf{60}, 104001 (1999).

\bibitem{Bal1}  V. Balasubramanian, J. deBoer and D. Minic, Phys. Rev. D
\textbf{65}, 123508 (2002); Klemm, Nucl. Phys. \textbf{B625}, 295
(2002).

\bibitem{GM1}  A. M. Ghezelbash and R. B. Mann, J. High Energy Phys. \textbf{%
01}, 005 (2002).

\bibitem{Deh1}  M. H. Dehghani, Phys. Rev. D \textbf{65}, 104003 (2002).

\bibitem{KLS}  P. Kraus, F. Larsen and R. Siebelink, Nucl. Phys. \textbf{B563%
}, 259 (1999).

\bibitem{DaM}  S. Das and R. B. Mann, J. High Energy Phys. \textbf{08}, 033
(2000); D. Birmingham and S. Mokhtari, Phys. Lett. \textbf{\
B508}, 365 (2001).

\bibitem{Deh2}  M. H. Dehghani, Phys. Rev. D \textbf{65}, 124002 (2002).

\bibitem{Deh3}  M. H. Dehghani, Phys. Rev. D \textbf{66}, 044006 (2002).

\bibitem{Deh4}  M. H. Dehghani and R. B. Mann, Phys. Rev. D \textbf{64},
044003 (2001).

\bibitem{Sken2}  S. de Haro, K. Skenderis and S. N. Solodukhin, Commun.
Math. Phys. \textbf{217}, 595 (2001); M. Bianchi, D. Z. Freedman
and K. Skenderis, Nucl. Phys. \textbf{B631}, 159 (2002); M.
Taylor-Robinson, hep-th/0002125.

\bibitem{Cal}  M. M. Caldarelli, G. Cognola, and D. Klemm, Class. Quantum
Grav. \textbf{17}, 339 (2000).

\bibitem{Tol}  R. C. Tolman, Phys. Rev. \textbf{35}, 904 (1930).

\bibitem{Crei}  J. D. Brown, J. Creighton and R. B. Mann, Phys. Rev. D
\textbf{50}, 6394 (1994).

\bibitem{Pec}  C. Peca and J. P. S. Lemos, Phys. Rev. D
\textbf{59}, 124007 (1999).

\bibitem{Cev}  M. Cvetic and S. S. Gubser, J. High Energy Phys. \textbf{04},
024 (1999).

\bibitem{Gub}  S. S. Gubser and I. Mitra J. High Energy Phys. \textbf{08},
018 (2001).

\bibitem{Davies}  P. C. Davies, Class. Quantum Grav. \textbf{6}, 1909 (1989).


\end{thebibliography}
\end{document}